\pdfoutput=1

\documentclass{svjour3}

\makeatletter
\def\cl@chapter{}
\makeatother

\usepackage[T1]{fontenc}
\usepackage[latin1]{inputenc}

\usepackage{dirtytalk}
\usepackage[detect-all]{siunitx}
\usepackage[inline]{enumitem}
\usepackage{listings}
\usepackage{booktabs}
\usepackage{graphicx}
\usepackage[most]{tcolorbox}

\usepackage{orcidlink}

\hypersetup{
colorlinks=true,
linkcolor=red,
citecolor=magenta,
urlcolor=blue}

\usepackage[capitalize,noabbrev]{cleveref}

\newtoggle{UseIEEETemplate}
\newtoggle{UseEMSETemplate}
\newtoggle{UseACMTemplate}

\togglefalse{UseIEEETemplate}
\toggletrue{UseEMSETemplate}
\togglefalse{UseACMTemplate}

\newif\ifACM

\iftoggle{UseACMTemplate}{
\acmConference[Short name]{Name}{Date}{Location}

\ACMtrue
}

\iftoggle{UseIEEETemplate}{
\usepackage[hyphens]{url}

\IEEEoverridecommandlockouts
\ACMfalse
}

\iftoggle{UseEMSETemplate}{
\ACMfalse

\usepackage{natbib}

\smartqed
}

\newcommand{\cor}{code review}
\newcommand{\cov}{code velocity}

\newcommand{\sloc}{\textsc{SLOC}}
\newcommand{\mcr}{{M}odern {C}ode {R}eview}

\newcommand{\oss}{open-source software}
\newcommand{\osscaps}{\textsc{OSS}}
\newcommand{\tta}{time-to-accept}
\newcommand{\ttfr}{time-to-first-response}
\newcommand{\ttm}{time-to-merge}

\newcommand{\dbsd}{{D}ragon{F}ly{BSD}}
\newcommand{\fbsd}{{F}ree{BSD}}
\newcommand{\nbsd}{{N}et{BSD}}
\newcommand{\obsd}{{O}pen{BSD}}

\newcommand{\SurveyStartDate}{September 14, 2022}
\newcommand{\SurveyFinishDate}{October 14, 2022}
\newcommand{\SurTotalResponses}{\num{110}}
\newcommand{\SurTotalFinishedSurveys}{\num{76}}
\newcommand{\SurTotalAgreedToConsent}{\num{75}}
\newcommand{\SurTotalAnonymousLink}{\num{50}}
\newcommand{\SurTotalSocialMedia}{\num{25}}
\newcommand{\SurTotalQuestions}{\num{16}}
\newcommand{\SurTotalRankingAnswers}{\num{65}}
\newcommand{\SurTotalRankingComments}{\num{3}}
\newcommand{\SurIndustryDevs}{\num{39}}
\newcommand{\SurOSSDevs}{\num{36}}
\newcommand{\SurOpenEndedResponses}{\num{47}}
\newcommand{\TotalLikertItems}{\num{24}}
\newcommand{\DifferingLikertItems}{\num{4}}
\newcommand{\MaxAcceptableMedianCrSize}{\num{800}}
\newcommand{\AppDevsCount}{\num{42}}
\newcommand{\SystemsDevsCount}{\num{20}}
\newcommand{\RealtimeDevsCount}{\num{3}}
\newcommand{\OtherDevsCount}{\num{8}}
\newcommand{\UnknownDevsCount}{\num{2}}
\newcommand{\PctOfDevsWhoAuthorCRs}{\num{89}}

\newcommand{\anonymize}[2]{#2}

\makeatletter
\def\observationcounter{\thetcb@cnt@observation}
\makeatother

\newtcolorbox[auto counter]{observation}{
	enhanced,
	breakable,
	colback=white!5!white,
	colframe=gray!75!black,
        coltitle=white,
        arc=.3mm, %
	title=\textbf{Observation~\thetcbcounter}
}

\makeatletter

\def\lst@makecaption{
  \def\@captype{table}
  \@makecaption
}

\makeatother
\makeatletter

\begin{document}

\title{Does Code Review Speed Matter for Practitioners?}

\iftoggle{UseEMSETemplate}{
\dedication{To all the unsung heroes who commit and review code every day.}
}

\iftoggle{UseACMTemplate}{
\author[1]{Gunnar Kudrjavets}
\orcid{0000-0003-3730-4692}
\affiliation[obeypunctuation=true]{
   \institution{University of Groningen\\}
   \city{Groningen, }
   \postcode{9712 CP}
   \country{Netherlands}}
\email{g.kudrjavets@rug.nl}

\author[2]{Ayushi Rastogi}
\orcid{0000-0002-0939-6887}
\affiliation[obeypunctuation=true]{
   \institution{University of Groningen\\}
   \city{Groningen, }
   \postcode{9712 CP}
   \country{Netherlands}}
\email{a.rastogi@rug.nl}

}

\ifACM

\fi

\iftoggle{UseIEEETemplate}{
\author{\anonymize{\IEEEauthorblockN{[Anonymous Authors]}}{
    \IEEEauthorblockN{Gunnar Kudrjavets}
    \IEEEauthorblockA{
        \textit{University of Groningen}\\
                9712 CP Groningen, Netherlands \\
                g.kudrjavets@rug.nl}

    \and

    \IEEEauthorblockN{Ayushi Rastogi}
    \IEEEauthorblockA{
        \textit{University of Groningen}\\
                9712 CP Groningen, Netherlands \\
                a.rastogi@rug.nl}
}}
}

\iftoggle{UseEMSETemplate}{
\author{Gunnar Kudrjavets \and
        Ayushi Rastogi
}

\titlerunning{Does Code Review Speed Matter for Practitioners?}

\institute{G. Kudrjavets~\orcidlink{0000-0003-3730-4692} \and
              A. Rastogi~\orcidlink{0000-0002-0939-6887} \at
              Bernoulli Institute for Mathematics, Computer Science and Artificial Intelligence\\
              University of Groningen,
              9712 CP Groningen, Netherlands\\
              Tel.: +31-50-363-3939\\
              Fax.: +31-50-363-3800\\
              \email{\{g.kudrjavets,a.rastogi\}@rug.nl}
}

\date{Received: date / Accepted: date}
}

\iftoggle{UseACMTemplate} {
\begin{abstract}

Increasing \cov\ is a common goal for a variety of software projects.
The efficiency of the \cor\ process significantly impacts how fast the code gets merged into the final product and reaches the customers.
We conducted a qualitative survey to study the \cov-related beliefs and practices in place.
We analyzed \SurTotalAgreedToConsent\ completed surveys from \SurIndustryDevs\ participants from the industry and \SurOSSDevs\ from the open-source community.
Our critical findings are
\begin{enumerate*}[label=(\alph*),before=\unskip{ }, itemjoin={{, }}, itemjoin*={{, and }}]
    \item
    the industry and open-source community hold a similar set of beliefs
    \item
    quick reaction time is of utmost importance and applies to the tooling infrastructure and the behavior of other engineers
    \item
    \ttm\ is the essential \cor\ metric to improve
    \item
    engineers are divided about the benefits of increased \cov\ for their career growth
    \item
    the controlled application of the commit-then-review model can increase \cov.
\end{enumerate*}
Our study supports the continued need to invest in and improve \cov\ regardless of the underlying organizational ecosystem.
\end{abstract}
}

\maketitle

\iftoggle{UseIEEETemplate}{
\begin{abstract}
\lipsum[1]
}

\iftoggle{UseEMSETemplate}{
\begin{abstract}
Increasing \cov\ is a common goal for a variety of software projects.
The efficiency of the \cor\ process significantly impacts how fast the code gets merged into the final product and reaches the customers.
We conducted a survey to study the \cov-related beliefs and practices in place.
We analyzed \SurTotalAgreedToConsent\ completed surveys from \SurIndustryDevs\ participants from the industry and \SurOSSDevs\ from the open-source community.
Our critical findings are
\begin{enumerate*}[label=(\alph*),before=\unskip{ }, itemjoin={{, }}, itemjoin*={{, and }}]
    \item
    the industry and open-source community hold a similar set of beliefs
    \item
    quick reaction time is of utmost importance and applies to the tooling infrastructure and the behavior of other engineers
    \item
    \ttm\ is the essential \cor\ metric to improve
    \item
    engineers have differing opinions about the benefits of increased \cov\ for their career growth
    \item
    the controlled application of the commit-then-review model can increase \cov.
\end{enumerate*}
Our study supports the continued need to invest in and improve \cov\ regardless of the underlying organizational ecosystem.
}

\iftoggle{UseEMSETemplate}{
} {}

\iftoggle{UseIEEETemplate}{
\end{abstract}
}

\iftoggle{UseEMSETemplate}{
\end{abstract}
}

\iftoggle{UseIEEETemplate}{
}

\section{Introduction}

Traditional software development methodologies, such as the waterfall model, focus on a rigid and highly predictable development and deployment schedule.
With the introduction of the Agile Manifesto in 2001, the focus of modern approaches to software development has shifted to continuous deployment of incremental code changes~\citep{martin_2002}.
As a result, Continuous Integration (\textsc{CI}) and Continuous Deployment (\textsc{CD})~\citep{fowler_continuous_2006} have become default practices for most of the projects in the industry and \oss.
A critical objective in the software industry is making code changes reach the production environment \emph{fast}.
The \cor\ process is a time-consuming part of evaluating the quality of code changes and approving their deployment.
With the wide adoption of \mcr~\citep{bacchelli_2013,sadowski_modern_2018} principles, engineers are now under more pressure than ever to review and deploy their code promptly.
One of the aspects of \mcr\ that negatively impacts software production is the perception that the \cor\ process is time-consuming~\citep{cunha_2021_2}.

In this paper, we define \emph{\cov} as \say{the time between making a code change and shipping the change to customers}~\citep{code_vel_definition_ast}.
We focus on the total duration of \cor\ completion and merge time.
We include a detailed description of related terminology in~\Cref{subsec:terminology}.

Based on our experiences in the last two decades with commercial and \oss\ development, we have witnessed various practices and beliefs related to increasing \cov.
The opinions related to increasing \cov\ range from the willingness to deploy code faster, even if it means increased defect density~\citep{kononenko_2016,facebook_philosophy_2011}, to taking as much time as necessary to \say{get the code right}~\citep{linux_lfd103}.
These opposing views raise questions about the developer community's prevalent attitudes and beliefs toward \cov.

Researchers have investigated the different aspects of the \mcr\ process in-depth~\citep{nazir_2020,weisgerber_small_2008,bacchelli_2013,czerwonka_2015}.
We are unaware of any studies focusing specifically on the beliefs, challenges, and trade-offs associated with increasing the \cov.
The primary goal of this study is to search for the beliefs and practices about \cov\ as-is and the context in which they hold.
We target a variety of experienced practitioners who contribute or review code for commercial and \oss.

We formulate the following research questions:

\begin{quote}
    \textbf{\textsc{RQ1}}: What beliefs and convictions are related to \cov\ for industry and \oss\ developers?
\end{quote}

\begin{quote}
    \textbf{\textsc{RQ2}}: What compromises are engineers willing to make to increase \cov?
\end{quote}

\begin{quote}
    \textbf{\textsc{RQ3}}: What are the essential suggestions from practitioners to increase \cov?
\end{quote}

To gather field data, we survey engineers who either submit or perform \cor s as part of their daily work.
We describe the recruitment of survey participants in~\Cref{subsec:survey_participants}.
Out of \SurTotalAgreedToConsent\ respondents, we classify \SurIndustryDevs\ individuals as industry participants and \SurOSSDevs\ as \oss\ contributors.
We asked survey participants various Likert-style questions related to the essence of our research inquiries.
Finally, we solicited free-form suggestions about how to increase \cov.

Our critical findings are that
\begin{enumerate*}[label=(\alph*),before=\unskip{ }, itemjoin={{, }}, itemjoin*={{, and }}]
    \item respondents working on both commercial and \oss\ respond similarly on Likert-type items (out of \TotalLikertItems\ items, only \DifferingLikertItems\ have a statistically significant difference between these two groups)
    \item while there is strong opposition to abandoning the \cor\ process, using the \emph{commit-then-review} model under some conditions can be acceptable
    \item respondents mainly focused on the development process, infrastructure and tooling support, response time, and the need to schedule pre-allocated time for \cor s to increase \cov.
\end{enumerate*}

Our study suggests that the maximum acceptable size of the \cor\ on the median is \MaxAcceptableMedianCrSize\ source lines of code (\sloc).
This number is an order of a magnitude larger than developer folklore and existing \cor\ guidelines suggest.
We find that the metric associated with \cor\ periods that engineers find the most useful is \ttm, followed by \tta\ and \ttfr.
That finding confirms what previous studies and grey literature have documented~\citep{izquierdo-cortazar_2017,tanna_2021}.

Issues of concern include
a need for more conviction that increased \cov\ benefits an engineer's career growth,
slow response times from either authors or \cor ers, and
faster fault detection from various parts of the infrastructure.

\section{Background and related work}

\subsection{Motivation for the study}
\label{subsec:motivation}

Developer velocity plays a significant role in software projects' success and the overall job satisfaction of developers.
The topic is important enough for Microsoft and GitHub to have a joint research initiative called Developer Velocity Lab~\citep{msft_dev_velocity_lab}.
According to Microsoft, \say{[i]mproving developer velocity is critical to continued satisfaction, iteration, and innovation in software teams}~\citep{mcmartin_introducing_2021}.
GitLab considers the reduction in the \cor\ time as the primary metric that describes the success of the \cor\ process~\citep{code_vel_gitlab}.
Data from Meta shows \say{a correlation between slow diff review times (P75) and engineer dissatisfaction}~\citep{riggs_move_2022}.
In this context, a \emph{diff} is a Meta-specific term equivalent to a \cor, pull request, or patch (used mainly in \oss).

In industry, the drive to increase the \cov\ is significant enough even to warrant the development of unique bots and tools.
These tools periodically remind either an author or reviewer that they block the completion of a \cor.
For example, at Microsoft, a bot that periodically \emph{nudges} developers \say{was able to reduce pull request resolution time by 60\% for \num{8500} pull requests,} with 73\% of these notifications being resolved as positive ~\citep{chandra_2022}.
Meta considers developer velocity as one of its critical investments during an economic downturn~\citep{vanian_internal_2022}.
The company gives engineers a general incentive to \say{[m]ove fast and break things}~\citep{facebook_philosophy_2011}.
In Meta's development philosophy, engineers expect certain defects to appear if it results in a faster product deployment~\citep{feitelson_2013}.
The startup culture practiced by many software companies encourages releasing new features \say{as fast as possible, for the sake of fuelling growth}~\citep{frenkel_2021}.

Another critical point in our inquiry is the differences in opinions about \cov\ between industry and \oss\ developers.
Fundamentally, the industry and \oss\ development process is motivated by different incentives.
However, the attitudes vary even in the context of \oss.
For the {L}inux kernel, \say{[t]he goal is to get the code right and not rush it in,} according to the official kernel development guide from the Linux Foundation~\citep{linux_lfd103}.
However, we notice the desire for increased \cov\ in Mozilla.
Based on a study about \cor\ practices in Mozilla, it is sometimes acceptable to be less thorough during \cor s if it speeds up the \cor\ process~\citep{kononenko_2016}.

Corporate policies are not necessarily dictated by what individual engineers think but by business needs.
Research shows that developers and managers have different views about productivity and quality~\citep{storey_2022}.
To discover the ground truth, we need to understand what engineers think is \say{right.}
The topic of \cov\ can surface strong emotions in engineers (\say{\dots\ I just hate it when things are unreviewed for days})~\citep{soderberg_2022}.
The survey mechanism that we use provides engineers with anonymity.
That anonymity enables engineers to freely share their opinions even if they contradict the company's or project's official policies related to \cov.

Empirical software engineering involves making daily trade-offs between various project characteristics.
The trade-off between increasing \cov\ and product quality has severe consequences because \say{poorly-reviewed code has a negative impact on software quality in large systems using modern reviewing tools}~\citep{mcintosh_2015}.
We want to study what attributes or values engineers are willing to compromise to achieve higher \cov.

\subsection{Terminology and metrics}
\label{subsec:terminology}

Most commercial organizations share a similar goal: \emph{reduce the duration of \cor s and consequently increase the \cov}.

The term \emph{\cov} can have different meanings depending on the context.
A customer-centric definition of \cov\ is \say{the time between making a code change and shipping the change to customers}~\citep{code_vel_definition_ast}.
As a quantifier characterizing code churn, it is defined \say{as the average number of commits per day for the past year of commit activity}~\citep{tsay_2017}.
In this paper, our definition focuses on the total duration of \cor\ completion and merge time.
We use \emph{\ttm} as the period from publishing the \cor\ to when accepted code changes are merged to the target branch~\citep{izquierdo-cortazar_2017}.
Terms like \emph{review time} (from publishing the patch until its acceptance)~\citep{tan_2019} and
\emph{resolve time}~\citep{zhu_2016} that is defined as \say{the time spent from submission to the final issue or pull request status operation (stopped at \emph{committed}, \emph{resolved}, \emph{merged}, \emph{closed}) of a contribution} are used as well.
Using the lifetime of a \cor\ coupled with merging time matches the period that the DevOps platforms such as GitLab optimize.
The formal definition for GitLab's metric is \say{duration from the first merge request version to merged}~\citep{code_vel_gitlab}.

Different commercial software companies measure various \cor\ periods.
Google's \cor\ guidance states that \say{\dots\ it is the response time that we are concerned with, as opposed to how long it takes a CL to get through the whole review and be submitted}~\citep{google_reviews_2022}.
The term \textsc{CL} in Google's nomenclature means \say{one self-contained change that has been submitted to version control or which is undergoing code review}~\citep{google_eng_practices}.
A study about \cov\ from Microsoft finds that critical points in time for engineers are \emph{the first comment} or \emph{sign-off} from a reviewer and when the \cor\ has been marked as \emph{completed}~\citep{bird_2015}.
A paper that investigates the performance of \cor\ in the {Xen} hypervisor project finds that \emph{time-to-merge} is the metric to optimize for~\citep{izquierdo-cortazar_2017}.
Anecdotal evidence from grey literature about the challenges in Capital One's \cor\ process presents a similar finding---\say{the most important metric was to calculate the \emph{cycle time}---that is, how long it takes from a PR being raised to it being merged (or closed)}~\citep{tanna_2021}.
Meta tracks a \emph{Time In Review} metric, defined as \say{a measure of how long a diff is waiting on review across all of its individual review cycles}~\citep{riggs_move_2022}.
One of the findings from Meta is that \say{[t]he longer someone's slowest 25 percent of diffs take to review, the less satisfied they were by their code review process.}

\subsection{Expectations related to \cov}

We find that \emph{expected \cor\ response times between industry and various \oss\ projects differ by order of magnitude}.
Google sets an expectation that \say{we expect feedback from a code review within 24 (working) hours}~\citep{winters_2020}.
Findings from Meta confirm that \say{reviews start to feel slow after they have been waiting for around 24 hour review}~\citep{chen_2022}.
Guidance from Palantir is that \say{code reviews need to be prompt (on the order of hours, not days)} and \say{[i]f you don't think you can complete a review in time, please let the committer know right away so they can find someone else}~\citep{palantir_code_reviews}.
Existing research into \cor\ practices at \textsc{AMD}, Google, and Microsoft similarly converges on \num{24} hours~\citep{rigby_2013}.
The published guidelines and studies match our industry experience of \num{24} hours being a de facto expected period for a response.

The requirements for \oss\ are less demanding than in the industry.
The \cor\ guidelines from the \textsc{LLVM} project~\citep{llvm_code_review} that focuses on various compiler and toolchain technologies set up an expectation that \say{code reviews will take longer than you might hope.}
Similarly, the expectations for Linux contributors are set to \say{it might take longer than a week to get a response} during busy times~\citep{linux_lfd103}.
The guidance for Mozilla is to \say{strive to answer in a couple of days, or at least under a week}~\citep{mozilla_cr_guidance_2021}.
For Blender, the expectation is that \say{[d]evelopers are expected to reply to patches [in] 3 working days}~\citep{blender_cr_guidance_2021}.

The etiquette for handling stale \cor s in \oss\ differs from the industry.
According to the guidelines from various projects, approaches like the Nudge bot are unacceptable~\citep{chandra_2022}.
The guidance for inactive \cor s for the Linux kernel is to \say{wait for a minimum of one week before requesting a response}~\citep{linux_lfd103}.
The \textsc{LLVM} project also recommends waiting for a week in case of inactivity before reminding the reviewers~\citep{llvm_code_review}.
The {FreeBSD} commit guidelines state clearly that
\say{the common courtesy ping rate is one week}~\citep{freebsd_code_review}.

\section{Methodology}

\subsection{Survey design}

The main goals of our survey are to collect data about the beliefs and experiences about \cov,
the trade-offs engineers are willing to make to increase the \cov, and suggestions from practitioners about how to increase it.
Our target audience was both commercial and \oss\ developers.

The survey consists of \SurTotalQuestions\ questions, with one additional asking for consent.
All the questions except the one that determined the participants' consent were \emph{optional}.
At the end of the survey, we asked participants if they wanted to share their contact information with researchers.
The first eight questions related to participants' demographics, such as experience with software development, \cor s, their role in the software development process, and the application domain they used to answer the survey questions.
This question block is followed by questions that ask participants to rank \ttfr, \tta, and \ttm\ in order of importance to optimize the \cov.
After that, we present four questions related to the benefits of \cov\ and potential compromises related to increasing \cov.
We inquire about the possibility of using either a post-commit review model~\citep{rigby_2006} or no \cor\ process.

Questions Q11, Q12, Q13, and Q14 contain multiple \emph{Likert-type items}~\citep{clason_1994}.
Because of the survey's limited number of questions, we do not use classic Likert scales.
Likert scales typically contain four or more Likert-type items combined to measure a single character or trait.
For example, researchers may have \num{4}--\num{5} questions about the timeliness of responses to a \cor.
Researchers then merge the results from these questions into a single composite score.
The categories we inquire about come from our combined subjective experiences with software development in the industry and open-source community.
Some choices, such as \say{Career growth} and \say{Job satisfaction,} are apparent.
Others, such as \say{Diversity, equity, and inclusion,} reflect the changing nature of societal processes.

The survey ends with asking participants about the maximum acceptable \cor\ size and waiting period, followed by an open-ended question about how the \cov\ can be improved.
We indicated to participants that the survey would take \num{5}--\num{7} minutes.
The complete list of survey questions is publicly accessible.\footnote{\protect\url{https://doi.org/10.5281/zenodo.7312098}}

\subsection{Survey participants}
\label{subsec:survey_participants}

\paragraph{Ethical considerations}
Ethical solicitation of participants for research that involves human subjects is challenging~\citep{felderer_2020}.
There is no clear consensus in the academic community about sampling strategy to solicit survey participants.
The topic is under active discussion~\citep{baltes_2016}.
In 2021, researchers from the University of Minnesota experimented with the Linux kernel~\citep{feitelson_2021}.
The approach the researchers used caused significant controversy and surfaced several issues surrounding research ethics on human subjects.
The fallout from the \say{hypocrite commits} experiment~\citep{wu_2021} forced us to approach recruiting the survey participants with extreme caution.

A typical approach is to use e-mail addresses mined from software repositories (e.g., identities of commit authors) to contact software developers.
Based on the recent guidance about ethics in data mining~\citep{gold_2021,jesus_m_mining_2020}, we decided not to do this.
While research related to \cor s has in the past utilized an existing relationship with a single specific developer community such as Mozilla~\citep{kononenko_2016} or Shopify~\citep{kononenko_2018}, we wanted to reach out to a broader audience.
We also wanted to avoid incentive-based recruitment mechanisms that offer rewards.
Existing findings suggest that monetary incentives increase survey response rates~\citep{smith_2019}.
However, they do not necessarily reduce the non-response bias~\citep{groves_2006}.
As a potential incentive, we promised that participants who share their contact information with us would be the first to receive the paper's preprint.

\paragraph{Recruitment strategy}
Our recruitment strategy used social media (Facebook, LinkedIn, Medium, Reddit, and Twitter) to spread the message to our network of professional software engineers, researchers, and students involved in the industry and \oss\ communities.
We started by composing a Medium post in a non-academic style, making the content reachable to a broader audience.
That post contained a link to our survey to make participation easier.
In addition, we contacted several individuals in commercial software companies and various \oss\ projects to ask their permission to share the survey invite with their developer community.
We received responses from Blender, Gerrit, \fbsd, and \nbsd.
These projects gave us explicit permission to post in a project's mailing list, or our contact person circulated the survey internally.

\paragraph{Survey summary statistics}
The survey was published on \SurveyStartDate, and closed on \SurveyFinishDate.
The survey system received a total of \SurTotalResponses\ responses.
Out of all the respondents, \SurTotalFinishedSurveys\ participants completed the survey, with \SurTotalAgreedToConsent\ agreeing to the consent form and answering the questions.
Of \SurTotalAgreedToConsent\ individuals who answered the questions, \SurTotalSocialMedia\ discovered the survey using social media and \SurTotalAnonymousLink\ via anonymous survey link.
For our analysis, we only used the surveys that participants fully completed.

\subsection{Survey data analysis}

The methods used to analyze data from Likert-style items are controversial without a clear scientific consensus~\citep{brown_2011,carifio_2007,chen_2020}.
This paper treats Likert-style items as ordinal measurements and uses descriptive statistics to analyze them~\citep{allen_2007,boone_2012}.
We do not treat ordinal values as metric because it can lead to errors~\citep{liddell_2018}.
For Likert-type items, we define three general categories:
\emph{negative} (\say{Strongly disagree,} \say{Disagree}),
\emph{neutral} (\say{Neither agree nor disagree,} \say{I don't know}), and
\emph{positive} (\say{Agree,} \say{Strongly agree}).
We added the choice of \say{I don't know} based on the feedback from the pilot tests that we used to refine the survey questions.

We use the content analysis to analyze the answers to Q17 (\say{In your opinion, how can code velocity be improved for your projects?}).
Two researchers independently manually coded all the responses to Q17.
Once the coding process was finished, the researchers compared their results and tried to achieve a consensus.
In case of disagreements, a third researcher acted as a referee.
Several themes and categories emerged as part of the open coding process.
We repeated the coding process till we classified all the responses under $7 \pm 2$ labels~\citep{miller_magical_1956}.

Numerical data, such as the maximum acceptable size of the \cor, was analyzed using custom code written in {R}.
Similarly, the statistical tests conducted in~\cref{subsec:RQ1} and \cref{subsec:RQ2} to evaluate the differences between various groups were implemented in {R}.

\section{Results}

\subsection{Demographics}
\label{subsec:demographics}

Most of the respondents to our survey are experienced software engineers.
We find that:

\begin{itemize}
    \item A \num{34}\% of respondents identified themselves as having \num{3}--\num{10} years of experience.
    \item A \num{62}\% stated that they have more than \num{10} years of experience.
\end{itemize}

Consequently, this experience translates to the time spent reviewing other people's code.

\begin{itemize}
    \item A \num{63}\% of survey participants have over \num{5} years of experience reviewing code.
    \item A \num{36}\% of respondents submit more than \num{10} \cor s per month,
and \num{43}\% submit \num{3}--\num{10} \cor s.
    \item A \num{54}\% of respondents conduct more than \num{10} \cor s per month
and \num{38}\% conduct \num{3}--\num{10} \cor s.
\end{itemize}

For the type of software the survey participants work on, \num{58}\% identified as developers working on application software such as mobile or desktop applications.
\num{27}\% of respondents stated that they work on systems software such as device drivers or kernel development.
Regarding different \cor\ environments, \num{95}\% of respondents use a code collaboration tool.
That tool may be public, such as {Gerrit} or {GitHub}, or the private instance of a company-specific tool, such as Google's Critique.
Nearly every respondent writes or reviews code as part of their role.
Only one individual stated that their role does not require reviewing code, and they only submit patches.
Out of the respondents, \PctOfDevsWhoAuthorCRs\% of developers author new code changes.
The rest of the survey participants have a different role, such as only reviewing the code.

\subsection{Grouping of respondents}

We divide our participants into two groups based on how they self-identify as a response to Q8 (\say{What type of software developer are you?}).
We use the following proxy to understand the difference between industry and \oss\ developers.
If a participant chose only \say{I work on closed-source and get paid} as a response, we classify them as \say{Industry.}
If one of the choices by participants was either \say{I work on open-source and get paid} or \say{I work on open-source and do not get paid,} then we classify them as \say{\osscaps.}
Based on that division, we ended up with \SurIndustryDevs\ participants from the industry and \SurOSSDevs\ respondents from \oss.

In Q9 (\say{Choose an application domain you are most experienced with for the remaining questions?}), we asked participants what type of software is their primary area of expertise.
We chose not to divide participants based on the abstraction level of the software.
We base that decision on the number of respondents and the varying size of the different groups.
Out of \SurTotalAgreedToConsent\ respondents, \AppDevsCount\ identified as someone working on application software,
\SystemsDevsCount\ on systems software,
\RealtimeDevsCount\ on real-time or critical software,
\OtherDevsCount\ on other types of software, and
\UnknownDevsCount\ chose not to answer the question.

\subsection{RQ1: beliefs and convictions related to \cov}
\label{subsec:RQ1}

\subsubsection{Expectations for the size and velocity of \cor s}

Our industry experience related to \cor\ size is highly variable.
We have participated in projects where \cor s that contained thousands of \sloc\ were standard.
Similarly, we have experience with projects where engineers required that authors split any review more extensive than \num{20}--\num{30} \sloc\ into separate reviews.
Existing research suggests that the size of \cor s impacts their quality and speed~\citep{jiang_2012,weisgerber_small_2008}.
Based on these findings, we investigate what engineers consider a maximum \emph{acceptable} size for a \cor.

Most \oss\ projects do not have fixed guidelines for an upper bound for a \cor.
Very little quantitative guidance exists for the size of \cor s.
Most guidelines use qualifiers such as \say{isolated}~\citep{llvm_patch_guidance}, \say{reasonable}~\citep{chromium_patch_guidance}, and \say{small}~\citep{postgresql_patch_guidance,phabricator_patch_guidance,macleod_code_2018}.
For stable releases of the Linux kernel, the guidance is \say{[i]t cannot be bigger than \num{100} lines \dots}~\citep{linux_no_more_than_100}.
Google engineering practices specify some bounds: \say{\num{100} lines is usually a reasonable size for a CL, and \num{1000} lines is usually too large}~\citep{google_eng_practices}.
The acronym \textsc{CL} means \say{one self-contained change that has been submitted to version control or which is undergoing code review}~\citep{google_eng_practices}.
As anecdotal evidence, a respondent to a survey about \cor\ practices states that \say{[a]nything more than \num{50} lines of changes, and my brain doesn't have the capacity to do a good code review}~\citep{alami_2020}.

The sentiment about larger patch sizes is generally negative.
A paper that investigates the efficiency of a \cor\ process finds that \say{patch size negatively affects all outcomes of code review that we consider as an indication of effectiveness}~\citep{santos_2017}.
The existing research directs developers towards more minor code changes.
Anecdotal assessment from the Chromium contributor's guide is that \say{[r]eview time often increases exponentially with patch size}~\citep{chromium_patch_guidance}.
A study about \cor\ performance finds that \say{review effectiveness is higher for smaller code changes}~\citep{baum_2019}.
Another study about participation in \mcr\ finds that patches with smaller sizes receive fewer comments than the larger patches, and larger patches go through more iterations~\citep{thongtanunam_2017,baysal_2015}.

\begin{figure}[!htbp]
    \centering
    \begin{minipage}{0.5\columnwidth}
        \centering
        \includegraphics[width=0.9\textwidth]{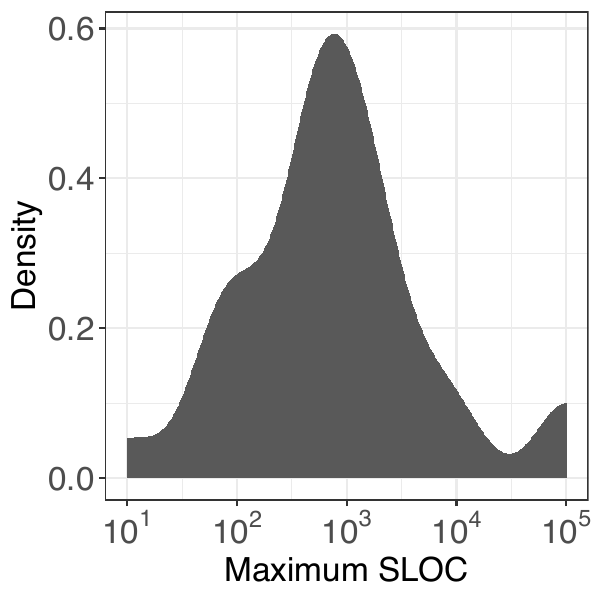}
        \caption{Source lines of code.}
        \label{fig:max_acceptable_sloc}
    \end{minipage}\hfill
    \begin{minipage}{0.5\columnwidth}
        \centering
        \includegraphics[width=0.9\textwidth]{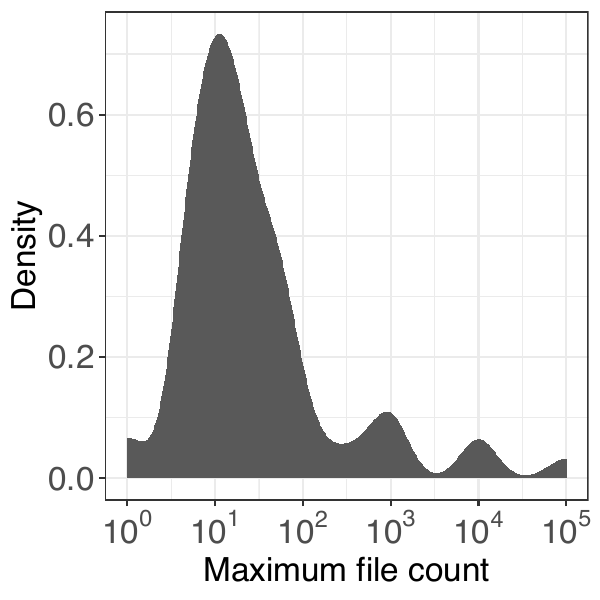}
        \caption{File count.}
        \label{fig:max_acceptable_file_count}
    \end{minipage}
\end{figure}

We asked study participants about the maximum acceptable number of \sloc\ and the number of files in the \cor.
\Cref{fig:max_acceptable_sloc} and \Cref{fig:max_acceptable_file_count} display the density plots (\say{smoothed histograms}) of both variables.
Before the analysis, we cleaned the data and removed entries that were \say{inconsistent with the remainder of the set of data}~\citep{gladitz_barnett_1988} and \say{surprising or discrepant to the investigator}~\citep{beckman_outlier_1983}.
We found only one entry for each metric that we considered an outlier.
Shapiro-Wilk tests~\citep{shapiro} confirmed that neither \sloc\ ($W = 0.33, p < .001$) nor the file count ($W = 0.15, p < .001$) were normally distributed.
A Mann-Whitney \textit{U} test~\citep{mann_whitney} indicated that the difference between medians for \sloc\ was not statistically significant $U(N_\mathrm{Industry} = \num{33}, N_\mathrm{\textsc{OSS}} = \num{26}) = \num{979.5}, z = \num{-0.16}, p = .87$.
Similarly, we do not observe differences for the number of files $U(N_\mathrm{Industry} = \num{34}, N_\mathrm{\textsc{OSS}} = \num{26}) = \num{1088.5}, z = \num{0.78}, p = .44$.

\begin{observation}
The median maximum acceptable number of \sloc\ for \cor\ is \num{800}.
This finding is surprising given the constant theme of suggesting that developers should aim for more minor changes.
\end{observation}

We discuss different \cor\ periods and how there is no consensus on what to optimize in~\Cref{subsec:motivation}.
Our goal is to understand what a heterogeneous collection of practitioners values the most when optimizing \cov.
We asked the participants to rank \ttfr, \tta, and \ttm\ in the order of importance.
A total of \SurTotalRankingAnswers\ participants ranked different \cor\ periods.
As a feedback, we also received \SurTotalRankingComments\ comments.
We present the results in~\Cref{tab:cor_period_rankings}.

\begin{table}[ht]
  \caption{Rankings of different \cor\ periods in the order of importance to optimize for \cov.
  Each column describes how many times a particular metric was ranked as a specific priority.
  The number of respondents and percentage of total responses per each entry is given.
  \textsc{TTFR} = \ttfr, \textsc{TTA} = \tta, \textsc{TTM} = \ttm.}
  \centering
  \label{tab:cor_period_rankings}
  \begin{tabular}{lrlrlr}
    \toprule
    \multicolumn{2}{c}{1\textsuperscript{st} priority}&
    \multicolumn{2}{c}{2\textsuperscript{nd} priority}&
    \multicolumn{2}{c}{3\textsuperscript{rd} priority}\\
    \midrule
1. \textbf{\textsc{TTM}}  & \num{32} (\num{49}\%) &1. \textbf{\textsc{TTA}} & \num{33} (\num{50}\%) &1. \textbf{\textsc{TTFR}} & \num{24} (\num{36}\%)\\
2. \textsc{TTFR}  & \num{19} (\num{29}\%) &2. \textsc{TTFR} & \num{22} (\num{34}\%) &2. \textsc{TTM} & \num{21} (\num{32}\%)\\
3. \textsc{TTA}  & \num{12} (\num{18}\%) &3. \textsc{TTM} & \num{10} (\num{15}\%) &3. \textsc{TTA} & \num{20} (\num{31}\%)\\
4. Other & \num{2} (\num{3}\%) &4. Other &\num{0} (\num{0}\%) &4. Other &\num{0} (\num{0}\%) \\
    \bottomrule
  \end{tabular}
\end{table}

In \num{49}\% of cases, the \ttm\ was ranked as the first priority metric to optimize.
This result is like the findings from the {Xen} hypervisor project study~\citep{izquierdo-cortazar_2017}, anecdotal evidence from the industry~\citep{tanna_2021}, and our personal  experiences for more than two decades.
One participant pointed out that \say{all of those metrics are totally irrelevant \dots} but did not clarify what else may be relevant.
Two other comments suggested a different set of metrics: \say{response time on changes in review by both author and reviewers} (like \emph{Time In Review} that Meta measures~\citep{riggs_move_2022}) and \say{[t]ime to re-review.}

\subsubsection{Perceived benefits of increased \cov}
\label{subsubsec:benefits_of_cov}

\begin{figure}[!htbp]
    \centering
    \includegraphics[width=0.8\textwidth,keepaspectratio]{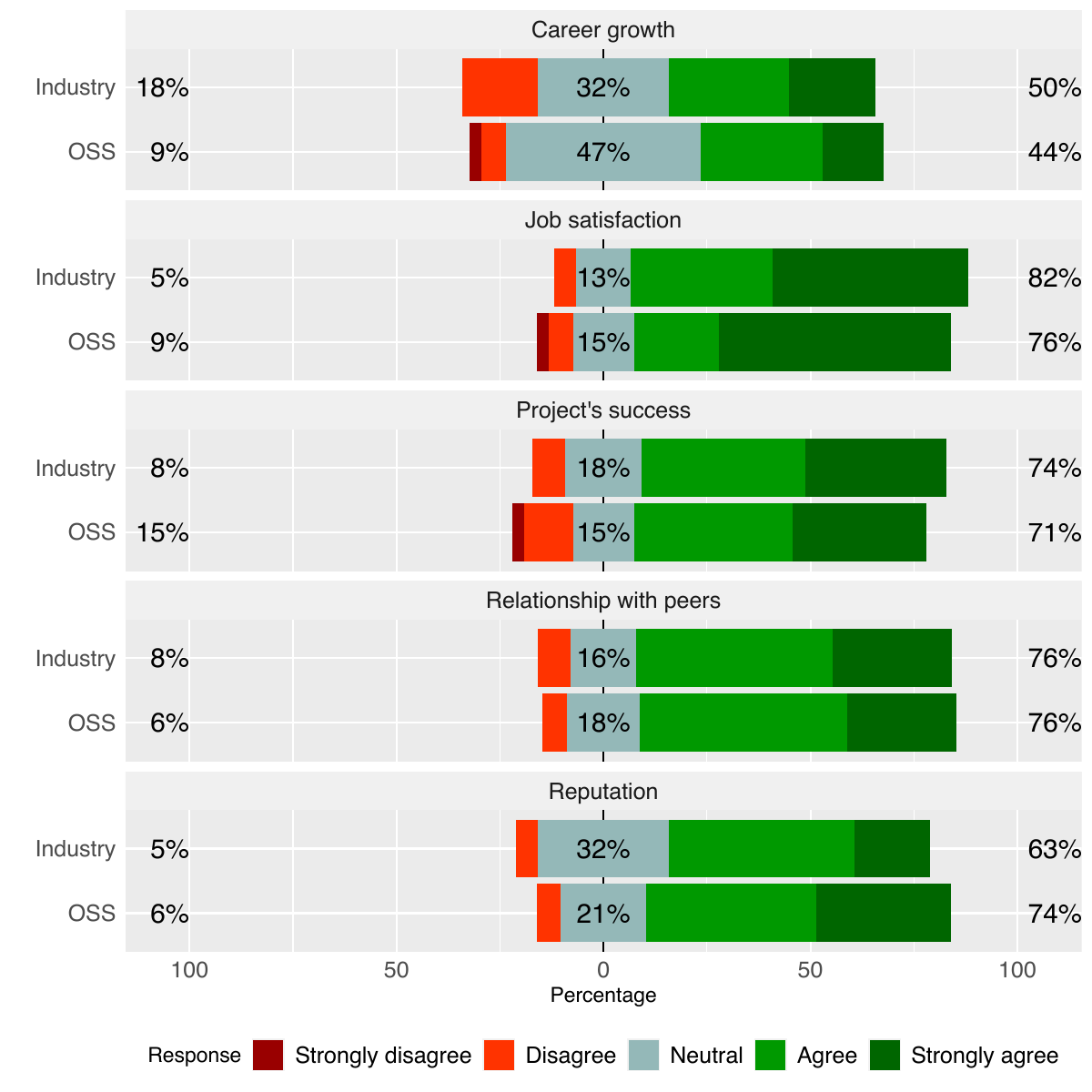}
    \caption{Likert scales for the Q11 (``Velocity of your code improves your \dots'').}
    \label{fig:Q11}
\end{figure}

We compare each Likert-style item between two groups separately using a Mann-Whitney \textit{U} test.
No statistically significant differences exist for any of the items between the industry and \osscaps\ groups.
\Cref{fig:Q11} shows that for most categories, participants perceive increased \cov\ as beneficial.
Participants think that \cov\ benefits aspects such as job satisfaction, reputation, or relationship with peers

The benefits of a \cor\ process generally have been associated with career development and growth~\citep{cunha_2021_1}.
However, in our study, the item with the lowest score is \say{Career growth,} where only \num{50}\% of the industry and \num{44}\% of \osscaps\ group respondents rated it positively.
Career growth in the corporate environment generally means increased professional scope, monetary rewards, and promotion speed.
This finding is somewhat concerning.
\emph{While intrinsic motivation is essential, it is hard to motivate engineers to conduct efficient \cor s if there is no substantial payoff}.

For the \osscaps\ group, one possible explanation is that it is much more challenging to define career progression in the \oss\ community than in a traditional corporate environment.
However, given that only \num{50}\% of responses from the industry rated the \say{Career growth} category positively, we think this topic is worth exploring further.

\begin{observation}
On a median, \num{74}\% of rankings are positive regarding the belief that increased \cov\ improves job satisfaction, project success, relationship with their peers, and reputation of engineers.
\end{observation}

\subsection{RQ2: compromises that are acceptable to increase \cov}
\label{subsec:RQ2}

\subsubsection{Commit-then-review model}

The foundation of the \mcr\ is the \emph{review-then-commit} model.
Some projects use the opposite of that approach.
One major software project that started using the \emph{commit-then-review} model was Apache~\citep{rigby_2008}.
While most of the projects have abandoned commit-then-review for review-then-commit, we wanted to study what developers think about the \emph{resurrection of the commit-then-review model}.
Several data points influence the decision to research this possibility.
We discuss them below.

\paragraph{Industry experience}
The primary motivation to survey developers about commit-then-review is our industry experience.
We have witnessed several complaints and discussions about the \say{slowness} of the review-then-commit model.
Developers are frustrated that even for trivial changes, such as one or two lines of code that fix formatting issues or compiler warnings, they must wait hours or days for someone to approve the code changes formally.
Even organizations that use cutting-edge approaches to software development, such as Meta, state that \say{every diff must be reviewed, without exception}~\citep{riggs_move_2022}.
We frequently observe this frustration about the inflexibility of the \cor\ process, primarily in teams of experienced and senior engineers.

\paragraph{Efficacy of \cor s}
Data from Microsoft shows that \say{[o]nly about \num{15}\% of comments provided by reviewers indicate a possible defect, much less a blocking defect}~\citep{czerwonka_2015}.
Given this finding, the trade-off between blocking the commits until the \cor\ finishes to satisfy the process versus optimizing for \cov\ and taking some risks needs investigation.
Our observations indicate that \emph{senior engineers consider \cor s to be efficient only if the reviewer is as or more senior than the author}.
The quality of various linters and static analysis tools has improved over the years.
We observe that tools automatically flag issues related to formatting, coding conventions, and fundamental coding issues without any human intervention.

\paragraph{Development process}
The establishment of \textsc{CI} and \textsc{CD} as de facto approaches to developing modern software shows that industry and \oss\ value the speed of software delivery as the critical success metric.

\begin{figure}[!htbp]
    \centering
    \includegraphics[width=0.8\textwidth,keepaspectratio]{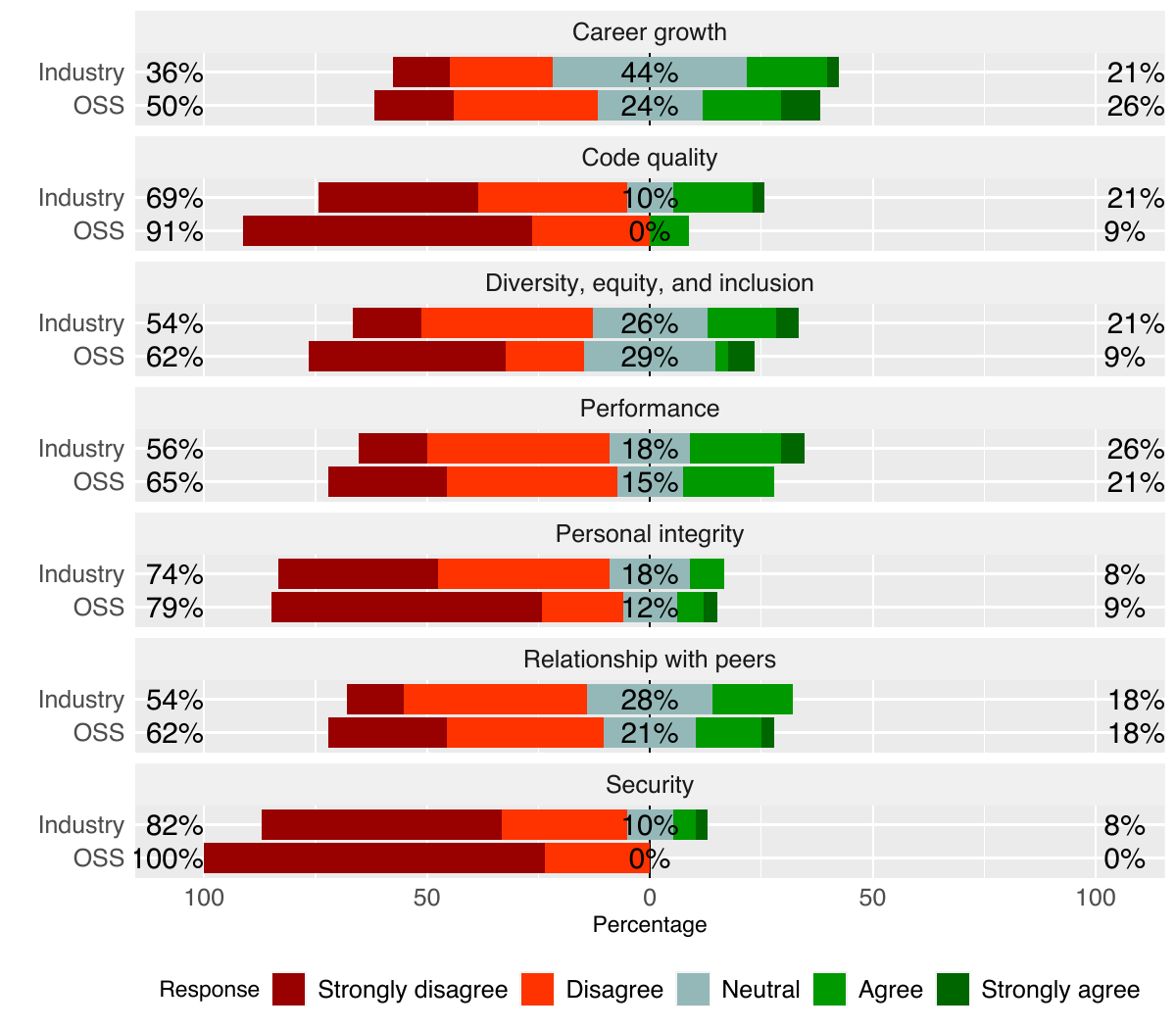}
    \caption{Likert scales for the Q12 (``I am willing to compromise on \dots\ if it improves code velocity'').}
    \label{fig:Q12}
\end{figure}

\Cref{fig:Q12} displays how much participants are willing to compromise on various characteristics to increase the \cov.
We compared each Likert-style item separately using a Mann-Whitney \textit{U} test.
There are significant differences only between two items: \say{Code quality} and \say{Security.}
A Mann-Whitney \textit{U} test for \say{Code quality} indicated that the difference in mean ranks is statistically significant $U(N_\mathrm{Industry} = \num{39}, N_\mathrm{\textsc{OSS}} = \num{34}) = \num{1664}, z = \num{2.65}, p = .008$.
Similarly, for \say{Security,} a Mann-Whitney \textit{U} test indicated that the difference in mean ranks is statistically significant $U(N_\mathrm{Industry} = \num{39}, N_\mathrm{\textsc{OSS}} = \num{34}) = \num{1621}, z = \num{2.33}, p = .02$.

An encouraging finding in our survey is what developers think about compromises related to code quality and software security.

\begin{observation}
For \num{100}\% of \osscaps\ developers, software security is something they are unwilling to compromise on to increase \cov.
\end{observation}

The \num{82}\% of industry developers who gave negative responses and \num{10}\% who gave neutral responses to this question share a similar sentiment.
Code quality is something that \num{91}\% of \osscaps\ developers and \num{69}\% of industry participants are not willing to negotiate over.
One potential explanation for the differences is that industry developers view code quality and security as one software characteristic regarding what they can make trade-offs.
On the other hand, the \osscaps\ developers are \say{true believers} who are not willing to compromise to release less secure software.

\begin{figure}[!htbp]
    \centering
    \includegraphics[width=0.8\textwidth,keepaspectratio]{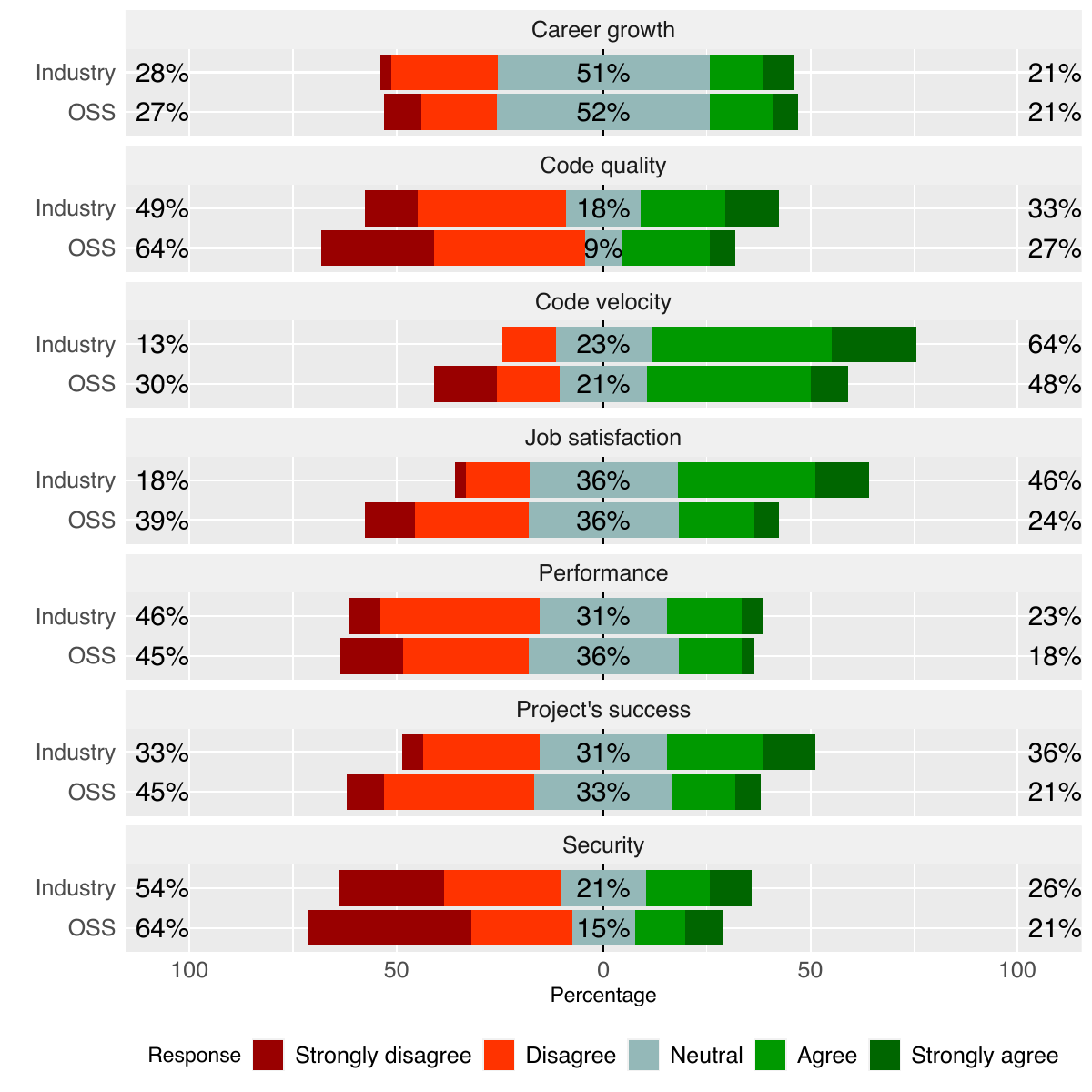}
    \caption{Likert scales for the Q13 (``I think the post-commit review model (changes are first committed and then reviewed at some point later) can improve \dots'').}
    \label{fig:Q13}
\end{figure}

As a next step, we asked participants what aspects of software a commit-then-review model can improve.
We display the results in~\Cref{fig:Q13}.
We compared each Likert-style item separately using a Mann-Whitney \textit{U} test.
There are significant differences only between two items: \say{Code velocity} and  \say{Job satisfaction.}
A Mann-Whitney \textit{U} test for \say{Code velocity} indicated that the difference in mean ranks is statistically significant $U(N_\mathrm{Industry} = \num{39}, N_\mathrm{\textsc{OSS}} = \num{33}) = \num{1603.5}, z = \num{2.12}, p = .034$.
The \cov\ is also a category that most participants thought could be improved.
Of industry respondents, \num{64}\% gave a positive response, with \num{48}\% of \osscaps\ respondents feeling similarly.
Application of the commit-then-review model means that developers no longer have to wait for a \cor\ to be complete.
The potential improvements in \cov\ are a logical result of this process change.

A Mann-Whitney \textit{U} test for \say{Job satisfaction} indicated that the difference in mean ranks is statistically significant $U(N_\mathrm{Industry} = \num{39}, N_\mathrm{\textsc{OSS}} = \num{33}) = \num{1620}, z = \num{2.28}, p = .023$.
Of industry respondents, \num{46}\% gave a positive response, with \num{24}\% of \osscaps\ respondents feeling similarly.
Approximately half of the survey participants from the industry think that their job satisfaction could improve with the commit-then-review model.
The difference between the industry and \osscaps\ is as significant as two times.
This finding makes sense because of how the industry evaluates the performance of software engineers.
Based on our experience, the ability to complete the assigned tasks on time and reduce \cor\ idle time is directly associated with engineers' anxiety levels and productivity.

For \say{Career growth,} \num{51}--\num{52}\% of respondents in industry and \osscaps\ chose a neutral response.
This finding indicates a need for more clarity regarding the relationship between \cov\ and its direct impact on an individual's career.
While it may be beneficial for an organization or a project to be released on a faster cadence, there is not necessarily a significant reward for individuals responsible for that cadence.
This finding is like the data from~\Cref{subsubsec:benefits_of_cov}, indicating that developers do not view an increase in \cov\ as beneficial to their careers.

\subsubsection{Abandonment of \cor s in favor of \cov}

Code reviews are optional in some contexts.
For example, the \fbsd\ project defines a committer as \say{an individual with \emph{write access} to the \fbsd\ source code repository}~\citep{freebsd_foundation_obtaining_2022}.
While committers are \say{required to have any nontrivial changes reviewed by at least one other person before committing them to the tree}~\citep{mckusick_design_2015}, the definition of \emph{nontrivial} is open to interpretation.
Therefore, experienced developers can commit code changes without the review at will.
We have also witnessed multiple instances in the industry where a similar practice is employed.
To increase the \cov, senior developers with a significant contribution history to the project have ignored the official \cor\ process or accepted each other's changes immediately to \say{satisfy the process.}

\begin{figure}[!htbp]
    \centering
    \includegraphics[width=0.8\textwidth,keepaspectratio]{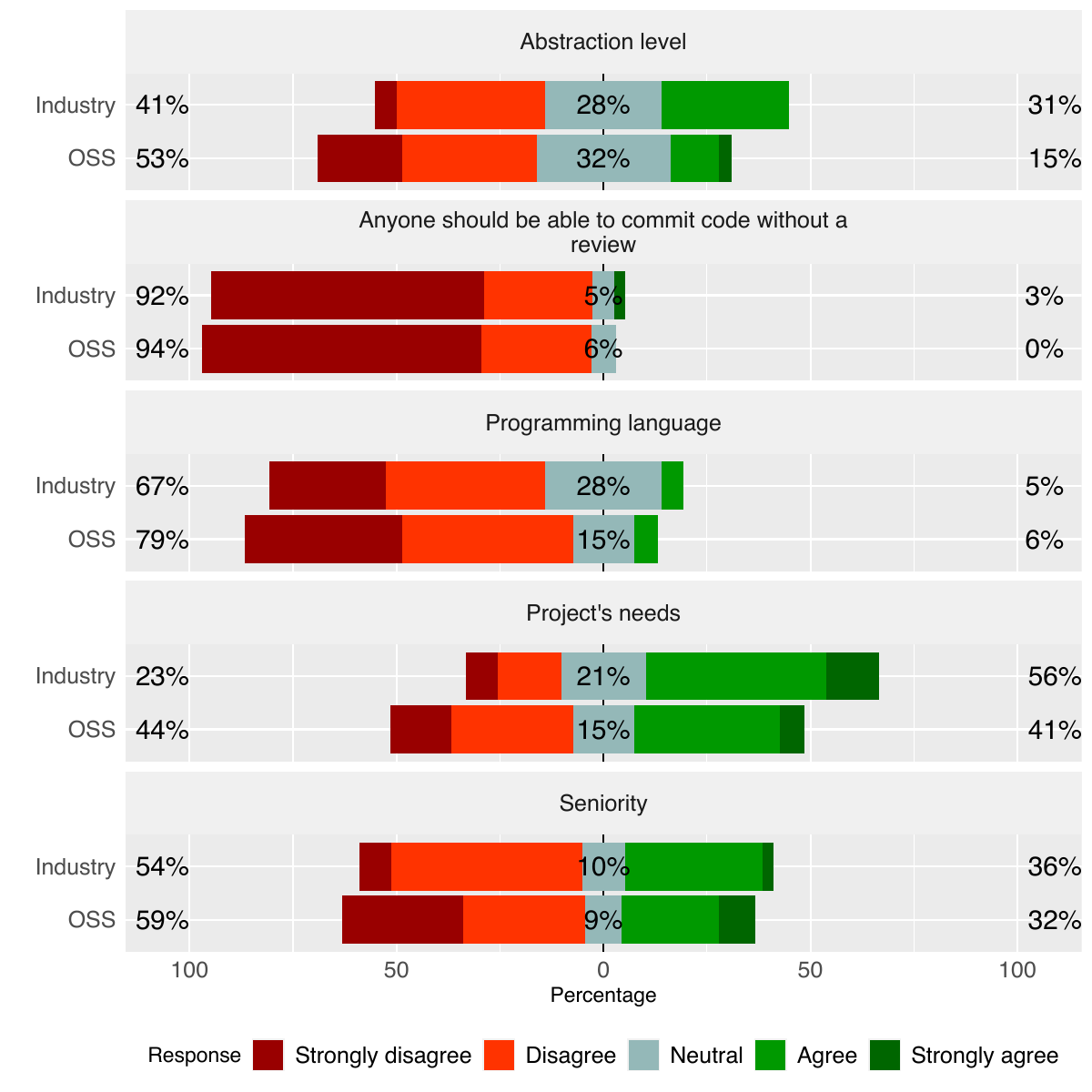}
    \caption{Likert scales for the Q14 (``Should engineers be allowed to commit their code without the code review depending on \dots'').}
    \label{fig:Q14}
\end{figure}

Based on the observations from the industry and the committer model used by projects such as \dbsd, \fbsd, \nbsd, and \obsd, we decided to survey what engineers think about the possibility of abandoning \cor s.
We asked participants under what conditions engineers can commit code without someone else reviewing that code.
We display the results in~\Cref{fig:Q14} and discuss them below.

We compared each Likert-style item separately using a Mann-Whitney \textit{U} test.
There were no statistically significant differences for any of the items between the \say{Industry} and \say{\osscaps} groups.

\begin{observation}
Developers are uniformly against abolishing the \cor\ process.
\num{92}\% of the industry and \num{94}\% of \osscaps\ respondents think the existence of the \cor\ process is valuable.
\end{observation}

The highest items with positive responses are \say{Project's needs} and \say{Seniority.}
For \say{Project's needs,} \num{56}\% of the industry and \num{41}\% of \osscaps\ respondents thought it permissible to commit code without conducting a \cor.
Based on our industry experience, this sounds reasonable.
Engineers must exercise their judgment in cases like build breaks or issues blocking an entire project and not blindly follow the process.
For example, suppose an application is not building, and an engineer has a potential fix.
In that case, it is reasonable to take a calculated risk to commit the changes immediately without waiting for hours for a \cor.

The choice of \say{Seniority} is reasonable as well.
Senior engineers typically have more tribal knowledge, related experience, and in-depth knowledge than junior engineers.
Therefore, if anyone can occasionally \say{break the rules,} it makes the most sense for them to do that.
In our industry experience, \cor s can find apparent mistakes.
However, finding problems in either complex algorithms or design nuances works best if a reviewer has a similar or higher level of knowledge.
Code reviews where a junior engineer reviews the senior engineer's code are effective in detecting defects only in a subset of cases.
Suppose the goal is to improve \cov.
In that case, we recommend that a project explicitly discuss the risk versus reward in a situation where senior engineers can exercise their judgment on when to require reviews for their changes.

\subsection{RQ3: suggestions to increase \cov}

\begin{figure}[!htbp]
    \centering
    \includegraphics[width=0.8\textwidth,keepaspectratio]{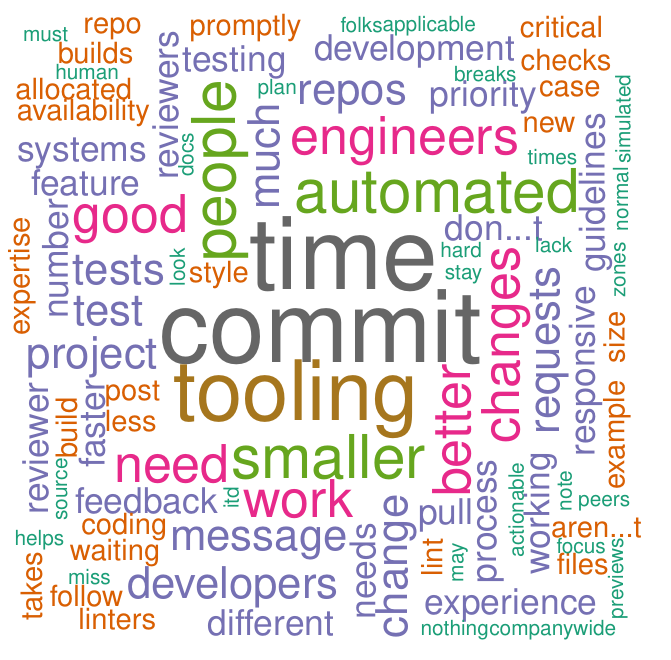}
    \caption{A word cloud of suggestions about how to improve \cov.}
    \label{fig:wordcloud}
\end{figure}

To solicit feedback from developers, we asked respondents, \say{In your opinion, how can code velocity be improved for your projects?}
We display the word cloud that summarizes the comments from the survey participants in~\Cref{fig:wordcloud}.
A word cloud is a widely used visualization technique to analyze and summarize qualitative data.
The rendering of the most frequently used words (e.g., \say{commit,} \say{time,} \say{tooling,} and \say{smaller}) causes them to be displayed more prominently.
While the word cloud is an initial indicator of the themes that emerge after analyzing the text corpus, a more detailed grouping of the results is necessary.

Two researchers manually analyzed and coded the \SurOpenEndedResponses\ comments received from the survey participants.
Each comment was assigned one or more labels depending on its content.
After the coding, researchers discussed the results and disagreements (less than ten), normalized the labels used for coding, and summarized the findings.
Our goal was to reach $7 \pm 2$ labels that adequately describe the nature of the suggestions~\citep{miller_magical_1956}.
We display the labels and their distribution in~\Cref{tab:survey-coding-labes}.

\begin{table}[ht]
  \caption{Different themes that result from coding the survey responses.}
  \centering
  \label{tab:survey-coding-labes}
  \begin{tabular}{lr}
    \toprule
    \textbf{Name}&\textbf{Count}\\
    \midrule
    Code review size & \num{5} \\
    Communication & \num{6} \\
    Coordination and scheduling & \num{13} \\
    Development process & \num{16} \\
    Early fault detection & \num{5} \\
    Infrastructure & \num{15} \\
    Response time & \num{12} \\
    \bottomrule
  \end{tabular}
\end{table}

\subsubsection{Improved conventions and standards}
We notice the desire for more formalized standards and the establishment of coding conventions.
Going forward, we use the notation R$i$ to indicate a respondent $i$.
An R5 states that \say{[e]stablishing and following company-wide standards for coding style and code review} can be helpful.
Similarly, R29 suggests that \say{[t]eam needs established coding conventions or tooling that enforces conventions to reduce debate.}
The standards help to set expectations and reduce the number of round-trips between an author and reviewer.
According to R11, it will be helpful to \say{decrease the number of surprises - have the review criteria prepared and explicit beforehand as much as possible/sensible.}
An R56 points out that
\say{[w]e need to improve our general code review guidelines - both for reviewer and reviewee (like "reviewer is not tester!").}
The sentiment is shared in R59 by asking for \say{stricter guidelines.}

\subsubsection{Prioritization of follow-up changes}
Determining clearly what is critical and what is not is another suggested improvement.
Feedback from R56 suggests that \say{code reviews shall be prioritised over new feature requests.}
An R30 suggests a potential two-phased approach to \cor s: \say{[b]e crisp on what is critical to fix and what can be done in a follow up} and \say{[s]top wasting time in LINT feedback and preferences about what code should look like and focus only on functionality and correctness.}
We have noticed similar behavior in the industry where reviewers try to separate the feedback into mandatory and optional.
The optional items are improvements that can be made in an independent \cor\ or later.

\subsubsection{Infrastructure and tooling improvements}

The critical requirement from infrastructure is fast capabilities for \emph{early fault detection}.
The general requirement is for \say{good code review tooling} for various tools \say{to perform more checks before actual "peer" review} (R37).
Developers want \say{[a]utomated testing, tooling} (R59).
The individual responses describe this need as
\say{[f]aster automated signals with more actionable error} (R6),
\say{[a]utomatic code linters, style cops, CI builds and CI test passes} (R9), and
\say{\dots automated CI/CD environments for initial checks, builds, tests \dots} (R28).

\subsubsection{Response time}

Compared to the status quo, the responses indicate the need for \emph{faster responses}.
The desire to respond quickly to \cor s is hardly a surprise.
Various existing studies and \cor\ guidelines specify that different periods of development processes should optimize~\citep{macleod_code_2018,google_reviews_2022,izquierdo-cortazar_2017}.
In addition to the increased anxiety caused by waiting for feedback, there are other drawbacks, such as the cost of a context switch and the potential for introducing new defects~\citep{czerwonka_2015}.
Respondents use phrases such as
\say{more responsive engineers} (R63),
\say{validation turn-around time} (R41),
\say{[c]ommunicating with reviewers promptly} (R1), and
\say{reducing the feedback loop} (R12)
to describe the potential improvements.
One of the respondents (R12) mentions that reducing \ttfr\ is crucial.
While tooling is essential, one respondent (R4) points out that
\say{the limiting factor is not really about tooling, but if reviewers are willing to spend the time it takes to review other people's changes.}

\subsubsection{Scheduling time for \cor s}

Existing research shows that developers spend \num{6.4} hours per week on reviewing code~\citep{bosu_2013}.
That is almost \num{20}\% of the typical \num{40}-hour work week.
In our industry experience, the management often classifies the time spent performing \cor s as the \say{cost of doing the business.}
Consequently, nobody accounts for this work during the formal planning process (if any).
The feedback from survey participants indicates that the time for conducting \cor s is an activity that planners need to include in the schedule formally.
Various responses demonstrate the need for better scheduling:
\say{have dedicated time to spend on code review} (R67),
\say{time allocated on their calendar to review the code} (R8), and\say{[m]ore time allocated for reviews} (R10).
Planning for the \cor\ time can
\begin{enumerate*}[label=(\alph*),before=\unskip{ }, itemjoin={{, }}, itemjoin*={{, and }}]
    \item reduce the potential anxiety developers have
    \item expose the cost of the \mcr\ process
    \item help to increase \cor\ quality because engineers can now take time to review code thoroughly.
\end{enumerate*}
We are not aware of any organizations in the industry that use the practice of accounting for a specific amount of \cor\ time.
According to R37, \say{commitment from engineers and the organization} can help to improve the \cov.

\subsubsection{Too much focus on speed}

An R33 provides an interesting observation: \say{[i]f anything development should be slowed down. Being in a (constant) hurry is a red flag.}
While we agree with this sentiment, we question if decreasing the \cov\ is possible for commercial software development.
We have rarely observed cases where industrial software projects attempt to decelerate the pace of code changes.
The rare situations in which that happens falls into two categories.
The first case is related to a stabilization period, such as weeks and months leading to shipping the product.
A second case results from the fallout from significant events, such as discovering repeated zero-day vulnerabilities in the code.

\subsubsection{Size of the \cor}

Other noteworthy themes include the \emph{size of the \cor} and \emph{communication}.
Requests such as
\say{[s]maller sized patches} (R1),
\say{smaller pieces to review} (R37),
\say{[s]maller merge requests} (R47),
\say{[h]ave less code to review} (R61), and
\say{[b]reaking up the use cases into smaller chunks} (R36)
indicate the desire for the upper bound of the \cor\ size.
While there is no definitive evidence to show that smaller sizes increase \cov,
the responses indicate that size is associated with overall job satisfaction.
\emph{The smaller \cor\ size is a strong preference amongst the engineers}.
Anecdotal guidance from {FreeBSD} suggests that \say{[t]he smaller your patch, the higher the probability that somebody will take a quick look at it}~\citep{freebsd_code_review}.

\section{Discussion}

We summarize the frequent themes that resulted from the analysis of the survey results.
The concrete application of suggestions and potential solutions to problems depends on the context, such as corporate culture or the project stage.

\paragraph{Commonalities between different groups}

The beliefs and willingness to make trade-offs are very similar between the practitioners in the industry and the \oss\ community.
Therefore, whatever solutions will manifest will be helpful for both groups.
The results from our survey indicate that for a majority of answers to Likert-style items, there are no statistically significant differences between industry and \oss\ developers.
The differences focus on career growth, job satisfaction, and what trade-offs engineers are willing to make to increase \cov.
The financial incentive structure is conceptually different between industry and largely unpaid \oss\ development.
We expected divergent views in these areas.

\paragraph{The need for speed}

Speed is the central theme across all the categories we cover in the survey.
Engineers expect the \cor\ process and the infrastructure to validate the code changes to be fast and responsive.
Most importantly, engineers need the \cor ers to promptly pay attention to the new code changes, follow-up questions, and any other open issues.
In our experience, some of these expectations and behavior are motivated by organizational culture.
Companies evaluate engineers' performance using metrics such as the number of open pull requests and their age, features deployed to the production environment, and even the \sloc\ that an engineer has committed.
These metrics can positively or negatively impact an engineer's career.
Therefore, it is reasonable to assume that engineers will focus on improving these metrics.

\paragraph{Engineer's career and \cov}
The impact of \cov\ on an engineer's career is unidirectional.
Actions such as missing deadlines,
not completing the work items on time,
or being unable to deploy the code in production to meet agreed-upon milestones negatively impact the engineer's career.
The item \say{Career growth} ranks lowest in items that increase in \cov\ impacts positively.
This finding is concerning at multiple levels.
Engineers can perform various actions to increase \cov.
For example, they can split their commits into isolated units,
write concise commit messages,
and establish an effective working relationship with their peers.
All these tasks are non-trivial and take time.
\emph{Based on the survey feedback, there is no objective payoff regarding career growth when engineers invest all that effort into increasing \cov}.

\paragraph{Splitting the code changes}

Previous research shows that patches introducing new features will receive slow feedback due to their size~\citep{thongtanunam_2017}.
There is no clear solution to mitigate this except splitting the code changes and sacrificing the cohesiveness of the code review.
The current trend in the industry is to split more prominent features into smaller ones and incrementally enable a subset of functionality.
However, implementation in small chunks is beneficial for only some features and products.
For example, the commonly used software Microsoft Office has \num{12000} feature flags~\citep{schroder_2022}.
Each feature flag can be enabled or disabled.
It is not immediately evident that the introduction of $2^n$ of different configurations software can operate under is beneficial for its maintenance.

\paragraph{Potential for the return of the commit-then-review model}
We recommend that projects consider the application of the commit-then-review model under some conditions.
 In situations where most engineers have noticeable experience,
 in-depth knowledge about the product,
 and can make informed trade-offs,
 letting developers use their judgment to decide when to ask for \cor s seems a good use of time.
A potential set of issues associated with this can be a decisions-process related to who is qualified enough to be permitted to act this way.
For example, can only engineers at a certain level make changes without a \cor?
That in itself may be divisive among the engineers.
Another issue could be engineers getting used to the \say{edit-compile-debug-commit-push} cycle and not requesting \cor s even for more extensive changes.
Industry can use a process that \oss\ uses to designate an individual as a committer~\citep{mckusick_design_2015}.

\section{Threats to validity}

Like any other study, the results we present in this paper are subject to specific categories of threats~\citep{shull_guide_2008}.

A primary threat to \emph{internal validity} is that a survey, due to its nature, is a self-report instrument.
We mitigate this threat by making the survey anonymous, indicating that all the questions are optional (except the consent), and avoiding any monetary incentives.

For \emph{conclusion validity}, we rely purely on what our participants reported.
Our primary data source is a collection of survey results.
We rely on the correct self-identification of survey participants to draw a subset of our conclusions.
To draw conclusions from our sample size and analyze the Likert-style items, we used non-parametric statistical methods recommended in the literature~\citep{allen_2007,boone_2012,mann_whitney,shapiro}.
Our survey could have reached a very homogeneous audience because we reached out to our contacts.
As a result, the views expressed may be like the ones that the authors of this paper hold.
We mitigated this concern by soliciting responses from Blender, Gerrit, \fbsd, and \nbsd\ developer communities.

For \emph{external validity}, the concern is if our findings are relevant in different contexts.
Because we decided not to solicit participants based on the data mined from various source code repositories, such as GitHub or Gerrit, we could not target specific demographics precisely.
We mitigated this by reaching out to several \oss\ projects and our connections in the industry to solicit responses.

\section{Conclusions and future work}

This paper presents the qualitative results from a survey about \cov\ and the beliefs and practices surrounding it.
We analyzed the responses to \SurTotalAgreedToConsent\ completed surveys.
Ethical solicitation of survey participants was a painstaking process requiring exhaustive usage of various social media channels.
Demographically, \SurIndustryDevs\ participants were from the industry, and \SurOSSDevs\ respondents were from the \oss\ community.
Based on what we know, this is the first paper that studies the trade-offs engineers make to increase the code velocity and critical impediments that block engineers from increasing code velocity even more.

The software development processes in the industry and open-source community have conceptual differences.
However, our survey suggests that most beliefs and trade-offs related to increasing \cov\ in these ecosystems are similar.
Engineers' critical concern is the payoff towards their career growth if code velocity improves.
A controlled application of the commit-then-review model scored the highest as a potential means to increase \cov.
Reduced software security is something that \num{100}\% of open-source and \num{82}\% of industry developers will not compromise, even if it means increased \cov.

In our future research, we plan to investigate the following topics:
\begin{enumerate*}[label=(\alph*),before=\unskip{ }, itemjoin={{, }}, itemjoin*={{, and }}]
    \item the selective application of the commit-then-review model in the industry
    \item the benefit of reward-based incentives to motivate engineers to react faster to \cor s
    \item the benefit of scheduling dedicated \cor\ time to achieve a more precise planning outcome.
\end{enumerate*}

\section*{Data availability}

The datasets generated and analyzed during the current study are available in the Zenodo repository.\footnote{\protect\url{https://doi.org/10.5281/zenodo.7312098}}
The dataset includes the anonymized responses to the survey and the {R} script that analyzes the survey data.

\iftoggle{UseIEEETemplate}{
\section*{Acknowledgments}
}

\iftoggle{UseACMTemplate}{
\section*{Acknowledgments}
}

\iftoggle{UseEMSETemplate}{
\begin{acknowledgements}
}

We thank all the survey participants for their insightful comments and suggestions.
We are incredibly grateful to the Blender, Gerrit, \fbsd, and \nbsd\ developer communities.
Our contacts in these teams either circulated the survey internally or allowed us to use their forum platforms or mailing lists to solicit survey participants.

\iftoggle{UseEMSETemplate}{
\end{acknowledgements}
}

\iftoggle{UseEMSETemplate}{
\section*{Conflict of interest}

The authors declare that they have no conflict of interest.
}

\iftoggle{UseACMTemplate}{
\bibliographystyle{ACM-Reference-Format}
}

\iftoggle{UseIEEETemplate}{
\bibliographystyle{IEEEtran}
}

\iftoggle{UseEMSETemplate}{
\bibliographystyle{spbasic}
}

\iftoggle{UseACMTemplate}{
\balance
}

\bibliography{survey}

\end{document}